\documentclass[12pt]{elsarticle}
\usepackage{hyperref}
\usepackage{graphicx}
\usepackage{dcolumn}
\usepackage{bm}
\usepackage{color}
\usepackage{version}
\usepackage{amsfonts}
\usepackage{soul}
\usepackage{float}
\usepackage{mathtools}

\begin{document}

\title{Minimalist Neural Networks training for Phase Classification in Diluted Ising Models}

\author{G. L. Garcia Pavioni}
\address{Facultad de Ciencias Exactas, Universidad Nacional de La Plata, La Plata, Argentina.}

\author{M. Arlego}
\address{IFLP - CONICET, Departamento de Física, Universidad Nacional de La Plata, C.C.\ 67, 1900 La Plata, Argentina.}
\address{Facultad de Ciencias Exactas, Universidad Nacional del Centro de la Provincia de Buenos Aires, Tandil, Argentina.}

\author{ C. A.\ Lamas}
\address{Facultad de Ciencias Exactas, Universidad Nacional de La Plata, La Plata, Argentina.}
\address{IFLP - CONICET, Departamento de Física, Universidad Nacional de La Plata, C.C.\ 67, 1900 La Plata, Argentina.}

\begin{abstract}
   In this article, we explore the potential of artificial neural networks, which are trained using an exceptionally simplified catalog of ideal configurations encompassing both order and disorder. We explore the generalisation power of these networks to classify phases in complex models that are far from the simplified training context. As a paradigmatic case, we analyse the order-disorder transition of the diluted Ising model on several two-dimensional crystalline lattices, which does not have an exact solution and presents challenges for most of the available analytical and numerical techniques. Quantitative agreement is obtained in the determination of transition temperatures and percolation densities, with comparatively much more expensive methods. These findings highlight the potential of minimalist training in neural networks to describe complex phenomena and have implications beyond condensed matter physics.
\end{abstract}

\maketitle

\section{Introduction}

In recent decades, machine learning and neural networks have experienced exponential growth in popularity and application across various fields of knowledge \cite{lecun2015deep, huang2023artificial, george2023review, chaka2023fourth, hirsch2023artificial}. This is largely due to the increasing availability of data in different domains and the ability to process them more rapidly and efficiently, thanks to the development of specialised software and hardware.

Artificial Neural Networks (ANN) are mathematical models composed by interconnected nodes, called neurons, that process information and learn through the identification of patterns in the data \cite{bengio2017deep}. Machine learning algorithms are based on the idea that computers can learn from data, being explicitly programmed for that purpose.

In the field of physics, machine learning methods and neural networks have been successfully employed to discover knowledge in complex systems \cite{mehta2019high}. These systems often comprise multiple interacting elements and exhibit emergent behaviours that cannot be predicted solely by considering the individual behaviour of each component \cite{anderson1972more}.

However, having data and resources abundance is not a guarantee of obtaining good results. In some cases, models trained with a large amount of data and tuning parameters (in the order of millions or more in the case of neural networks) can capture intrinsic details of the training data, a phenomenon known as `overfitting' \cite{ying2019overview}. The same can occur when there is a scarcity of data, where the model learns all the details of the available data, which, being scarce, lose representativeness of the overall set. This phenomenon can conflict with the main objective of neural networks: their ability to generalise \cite{kawaguchi2017generalization}. 

Generalisation refers to the network ability to capture knowledge that emerges from the training data and is also useful for describing other related contexts. The scale of the problem at hand is a determining factor for the amount of resources required for training, in order to achieve a balance between model accuracy and generalisation capacity.

It is generally accepted that the analysis of complex systems in physics requires significant resources. However, this study shows that, at least for certain types of complex systems found in condensed matter and other related applications, achieving a high degree of generalisation is possible even with deliberately limited resources.

In this work, we investigate the properties of the diluted Ising model on two-dimensional crystalline lattices, which is considered a  paradigmatic complex model in condensed matter physics. The diluted Ising model is a variant of the pure Ising model \cite{mccoy2014two}, widely employed in statistical physics to describe some magnetic materials. 
The Ising model exhibits a second-order transition from a disordered phase (paramagnetic) at high temperatures to an ordered phase (ferromagnetic) at low temperatures. In the transition, the system displays a variety of properties typical of complex systems, including long-range correlations, criticality, and scale invariance, which are distinctly different from the nature of the ferromagnetic and paramagnetic phases.

The diluted Ising model \cite{stauffer, Stinchcombe-Dilute-Ising} introduces an additional level of complexity to the Ising model by allowing the presence of vacancies. This model proves to be useful in investigating materials with impurities or structural defects, providing a simplified yet effective approach to describe the physics of magnetic systems in the presence of disorder.

A closely related phenomenon to the diluted Ising model is percolation \cite{stauffer}, which describes the propagation of a fluid through a porous medium, achieving percolation at a critical porosity. In the context of the diluted Ising model, the percolation transition occurs at a critical spin density, below which the system remains disordered, even at zero temperature. The interplay between dilution and temperature gives rise to a phase diagram where the transition between the ordered and disordered states takes place along a curve whose shape has been investigated using a variety of numerical methods and analytical approximations, as in references \cite{martins2007, p_cH, selke, Neda}, among others.

However, the main challenge emerges near the percolation density \cite{stauffer}, where singularities predicted by approximate methods arise, severely limiting the predictions of numerical techniques such as Montecarlo simulations. For this reason, the study of the phase transition in the diluted Ising model remains an active area of research, for which a variety of theoretical and numerical methods have been developed to achieve a better understanding of critical phenomena.
In this context, we will analyse the capacity of neural networks to determine the order-disorder transition in diluted Ising models on various two-dimensional crystalline lattices. 

It is worth noting that a variety of machine learning methods have been applied to the Ising models, as in references \cite{alexandrou2020critical, efthymiou2019super, d2020learning, scriva2023accelerating, zhang2021ising, walker2020deep, Potts2020, morningstar2017deep, acevedo-NNFlow, acevedo2021phase}, and to the percolation problem \cite{shen2021supervised, shen2022transfer, bayo2022machine, honecker2023percolating, YU2020125065, zhang2019machine, zhang2022machine}. Regarding percolation, these previous studies primarily investigate the geometric aspects of the problem and are therefore unrelated to our study.

Another distinction is that in this paper, we focus on the potential of what we refer to as `minimalist' networks, which efficiently generalize beyond the training context.

The concept of minimalist neural networks is broad, as illustrated in the references \cite{ho2022searching, kan2018minimalistic}. In this study, we specifically refer to a training strategy by using a minimal set of configurations that represent ideal ordered or disordered patterns to train a feed-forward neural networks, such as Dense Neural Networks (DNNs) with a single hidden layer. 

The objective of this approach is to develop a trained neural network equipped with a diverse and extensible catalog of ideal configurations, capable of representing the impact of various types of couplings within multiple crystalline lattices, encompassing extreme values at both zero and infinite temperatures. Through this process, the neural network effectively learns to classify these limiting configurations and subsequently extends its capabilities to handle more complex configurations.

How can we evaluate the generalisation capability of the trained minimalist neural networks? One way is to face the task of classifying configurations that were not encountered during training. To achieve this, we employ diluted ferromagnetic Ising models on two different two-dimensional crystalline lattices: the square lattice and the honeycomb lattice. These models contain two essential ingredients that are not present in the ideal training configurations: finite temperature and dilution (vacancies).

By evaluating the network performance on these classification tasks, we can determine its ability to generalise and apply the learned knowledge to new situations. 

To generate samples of the diluted Ising model at finite temperature and vacancy concentration (complementary to the spin concentration), we perform Montecarlo simulations \cite{Barkema}. Unlike typical usage, in this case, we do not use the simulations to train the neural network but rather to validate and test its generalisation capability. From this perspective, it is irrelevant whether the data comes from simulations or experiments since the network has already been trained on a set of ideal configurations and does not depend on these data for training, although it does rely on it for validation.

In summary, this work presents the use of minimalist neural networks in statistical physics, specifically in the diluted Ising model. Through Montecarlo simulations and other analytical approximations as benchmark, it is showed how these networks, trained with a catalogue of idealised configurations, can describe complex phenomena and accurately determine order-disorder phase transitions in different types of lattices and capture highly divergent behaviour in regions that pose challenges for other methods, with comparatively low computational cost.

The paper is structured as follows. In Section \ref{Models}, we provide a concise overview of the models and crystalline lattices examined in this study. This section begins with an exploration of the Ising model, followed by a description of the diluted Ising model and its relationship with the percolation phenomenon.

Section \ref{entrenamiento} delves into the process of generating the synthetic data catalog, describes the neural network's architecture, and how the network is trained using the cataloged data.

Section  \ref{results} is devoted to presenting the primary findings of the study, which are divided into three parts. Initially, we scrutinize the neural network's capacity for generalization when categorizing phases within the pure Ising model. This process is then extended to encompass the diluted Ising model, ultimately culminating in an analysis of percolation. Finally, the phase transition line is fine-tuned, utilizing the neural network's predictions on an approximate model to ascertain the percolation density. These parts are systematically explored for both square and honeycomb lattices.

The paper concludes with section \ref{conclusions} dedicated to summarizing the principal findings of the study. Additionally, to enhance the overall readability of the paper, technical details concerning the generation of synthetic Montecarlo data for model validation are presented in the \ref{appendix}.

\begin{figure} [t!]
    \centering
    \includegraphics[width=0.4\columnwidth]{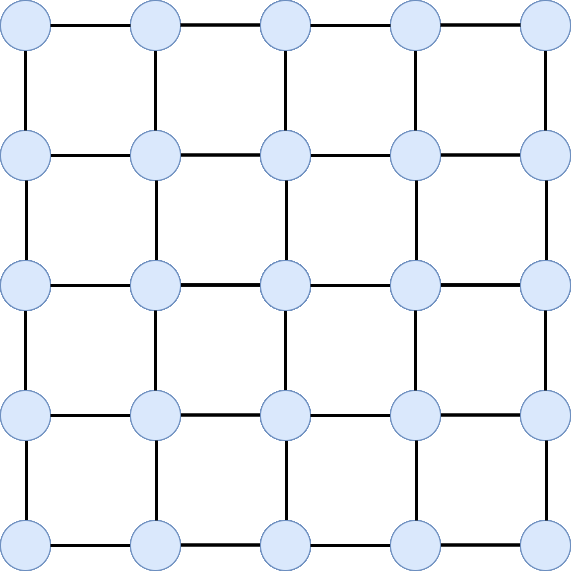}
    \includegraphics[width=0.4\columnwidth]{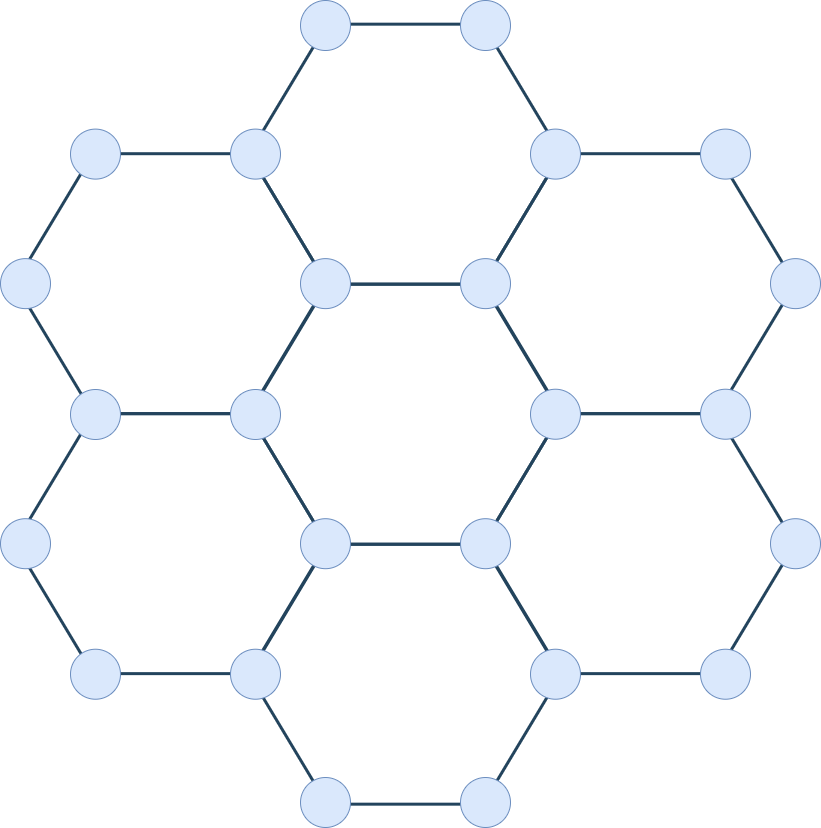}
    \caption{The lattices studied in this work. 
    The square lattice can be seen in the left panel, whereas the right panel illustrates the honeycomb lattice. It's important to note that in both scenarios, we have applied periodic boundary conditions.
    }
    \label{redes de espines}
\end{figure}

\section{Models and Lattices}
\label{Models}

In this section we provide a brief survey of the models and crystalline lattices analyzed in this paper. This section starts with an description of the Ising model, followed by the diluted Ising model and its connection to the percolation.

\subsection{Ising model}
 
In the field of solid-state physics, one of the most interesting phenomena is ferromagnetism \cite{simon2013oxford}. This phenomenon emerges in certain metals where a finite fraction of atomic spins spontaneously align in the same direction, thereby generating a macroscopic magnetic field. This aligned phase is commonly referred to as the ferromagnetic phase. The ferromagnetic phase occurs only when the temperature is below a critical value, denoted as $T_c$, also known as the Curie temperature. As the temperature increases beyond this critical value, the atomic spins become randomly oriented, resulting in a net zero magnetic field. This disordered phase is referred to as the paramagnetic phase.

Although ferromagnetism is fundamentally rooted in quantum mechanics, certain aspects of the phenomenon can be effectively described in classical terms, within the framework of the Ising model \cite{isin,walker2023student}. This is one of the most studied models in physics due to its exact solution in one and two dimensions, which makes it useful for testing various approximations and then comparing them with analytical solutions. In addition to its significance in magnetism, the Ising model serves as a paradigm for complex systems with two degrees of freedom. For applications and extensions of the Ising model in various fields, please refer to references \cite{aplicaciones_ising}.

The Ising model consists of an n-dimensional lattice comprising $N$ fixed points or `sites' each identified by a generic index \textit i. At each site, there is a spin variable denoted as $S_i$, representing an `up' (1) or `down' (-1) configuration. The entire system's configuration is characterized by the set of these spin values $\{S_i\}$. In absence of an external magnetic field, the model's Hamiltonian is defined as
\begin{equation}
        \mathcal{H}=-\sum_{ i,j } ^{} J_{i,j}\epsilon_{i} \epsilon_{j} S_{i}S_{j},
        \label{H}
\end{equation}
where the sum is extended to all sites $i, j$. The two-dimensional ($n=2$) lattices analyzed in this study are the square and honeycomb lattices, and are shown in the left and right panels of Figure \ref{redes de espines}, respectively.

In the case in which the interaction $J_{i,j} > 0$ the spins  align in order to minimize the energy, as defined by Hamiltonian $\ref{H}$, leading to a ferromagnetic behavior. Conversely, for $J_{i,j} <0$, the energy is minimized when neighboring spins at sites $i,j$ point in opposite directions, resulting in the formation of \textit{N\'eel} order. On the other hand, the term $\epsilon_{i}=1 (0)$ represents the presence (absence) of a spin at the $i$ site, making it possible to describe diluted Ising models. In this context, the spin density of the crystal lattice $\rho$ is defined as $\rho=\sum_{i=1}^N \epsilon_{i}/N$, where $N$ is the number of sites in the lattice. We use periodic boundary conditions to ensure that every site has an equal number of neighbors and maintains a consistent local geometry.

The pure ($\rho = 1$) 2D square lattice with a uniform interaction strength among first neighbors ($J_{i,j} = J$) exhibits an exact solution \cite{onsager1944crystal} and undergoes a second-order phase transition at the critical temperature $T_c = \frac{2|J|}{k_B \ln(1 + \sqrt{2})} \approx 2.269 \frac{|J|}{k_B}$, where $k_B$ represents Boltzmann's constant. For the pure honeycomb lattice, the critical temperature is also analytically known as $T_c = \frac{2|J|}{k_B \ln(2 + \sqrt{3})} \approx 1.519 \frac{|J|}{k_B}$. This transition occurs from a disordered state (paramagnetic phase) at temperatures $T > T_c$ to an ordered state (ferromagnetic phase for $J > 0$) at temperatures $T < T_c$.

In the 2D Ising model, the second-order phase transition is defined by an order parameter selected to be zero in the paramagnetic phase and non-zero in the ferromagnetic phase. In the ferromagnetic scenario, this order parameter is the spin magnetization per spin
\begin{equation}
        m =\frac{1}{N}\sum_{ i=1 } ^{N}  S_{i},
        \label{m}
\end{equation}

\subsection{Dilute Ising Model and percolation}

An important concept that will be discussed is the introduction of defects in a pure spin lattice. This involves the removal of spins from specific sites ($\epsilon_{i}=0$), consequently altering the density $\rho$ of the spin lattice. In this context, eq. \ref{H} gives rise to the diluted Ising Model. 

In the scenario of a dilute lattice ($\rho<1$), it is well-established that as the spin concentration $\rho$ decreases, the critical point ($T_c$) of the phase transition also decreases. This is to be expected, since the critical temperature depends on the interaction between neighboring spins, which decreases with the dilution.

\begin{figure}[t!]
    \centering
    \includegraphics[width=0.95\columnwidth]{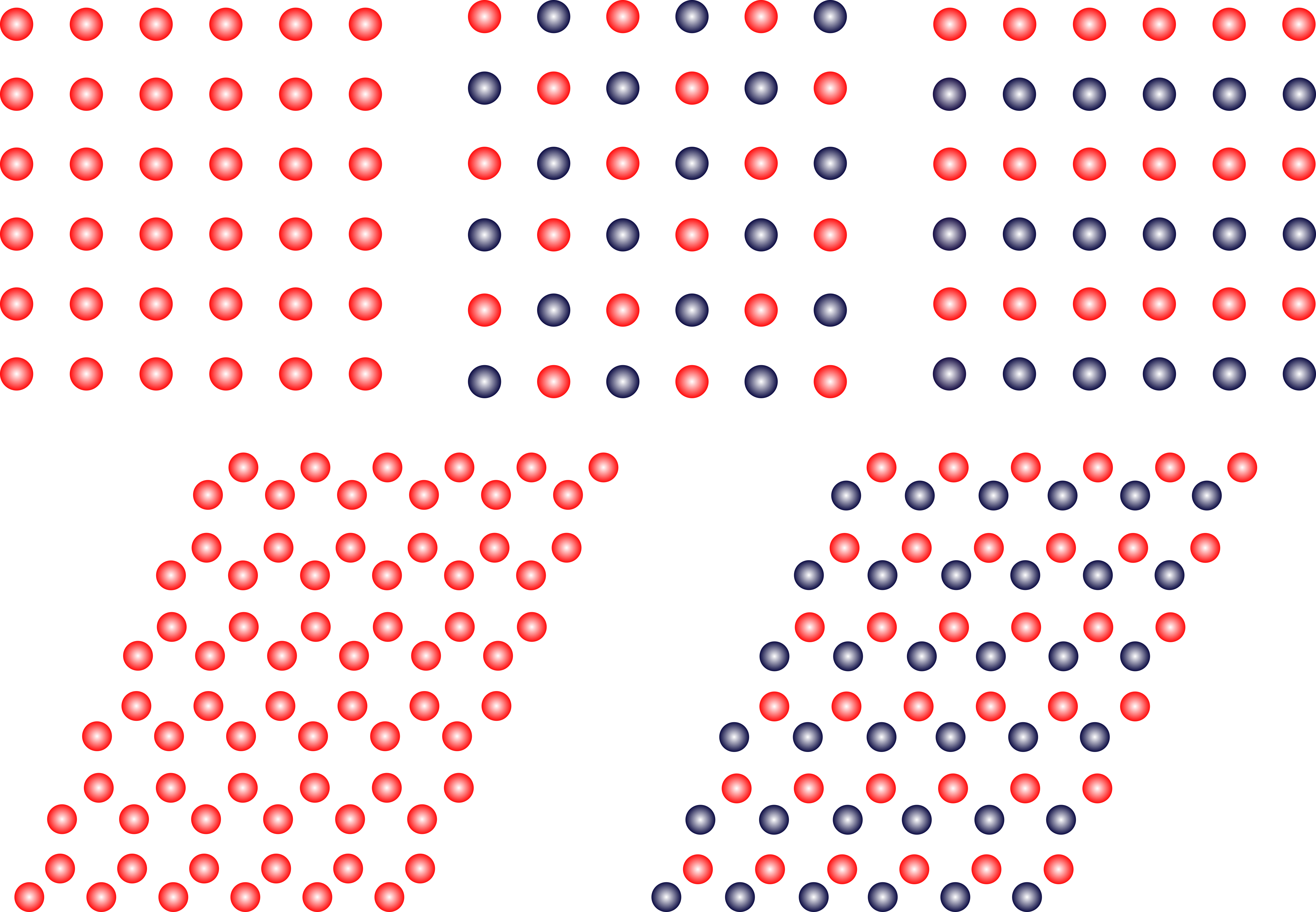}     
    \caption{Schematic representation of the configuration catalog utilized in the training process. Red dots indicate spin-up sites, while blue dots denote spin-down sites. The depicted configurations, presented from left to right, encompass ferromagnetic, antiferromagnetic (Néel), and stripe patterns. These configurations are showcased for both the square lattice (upper panel), and the honeycomb lattice (lower panel) for the first two configurations. Additionally, the catalog includes random configurations, representing the paramagnetic state at infinite temperature (not showed here).    
    }
    \label{fig:catálogo-square}
\end{figure}

The percolation density $\rho_c$ is defined as the density beyond which the system remains disordered at any temperature. Therefore, for values of $\rho \leq \rho_c$, the system remains in the paramagnetic phase regardless of the temperature ($T$). In other words, the critical temperature is zero, $T_c (\rho_c)=0$ \cite{stauffer}. The percolation densities, denoted as $\rho_c$, for the square and honeycomb lattices have been obtained approximately with high precision. Specifically, for the square lattice, $\rho_c \approx 0.59274621(13)$ \cite{p_c}, and for the honeycomb lattice, $\rho_c \approx 0.6962(6)$ \cite{p_cH}. It is important to mention that alternative definitions of percolation density exist, which emphasize the geometric aspects of this phenomenon and are unrelated to a transition temperature. However, these definitions fall outside the scope of the diluted Ising model and are not explored in this study.

The phase diagram of the dilute Ising model in the temperature-dilution plane encompasses the paramagnetic and ferromagnetic phases, divided by a critical transition line. In contrast to the pure case, the exact form of this critical line remains unknown. In the vicinity of the pure case, the transition exhibits a linear behavior, with the critical temperature decreasing as the dilution (i.e., the number of vacancies) increases. However, near a critical dilution value, this pattern abruptly changes, signifying the onset of percolation transition, where the system fails to order even at zero temperature. In this region, the transition curve presents a logarithmic divergence. Yet, directly capturing this behavior through Montecarlo simulations is a highly demanding task. However, several numerical and analytical investigations, as reported in references \cite{selke,Neda}, have proposed that the phase transition curve $T_c(\rho)$ satisfies the following relationship

\begin{equation}
    \frac{T_c(\rho)}{T_c(1)}= - \frac{K}{\log(\rho - \rho_c)},
    \label{curva_Tc_vs_p}
\end{equation}
where $K$ is a non-universal constant dependent on the crystal lattice.

\section{Minimalist Neural network training}
\label{entrenamiento}

In this section, we outline the process of creating the synthetic data catalog, the architecture of the neural networks employed, and describe their minimalist training approach utilizing the catalog data.

In typical machine learning phase classification problems, training data is derived from either experimental data or numerical simulations. This approach incurs significant costs and necessitates prior knowledge of the system's phases.

However, in this study, we adopt a distinct approach by employing a catalog of synthetic training data. By `synthetic', we mean that the spin configurations used for training are explicitly generated, with each lattice site assigned a spin value of either $+1$ or $-1$, depending on the desired spin configuration. These configurations represent ideal ordered states at absolute zero temperature and ideal disordered, or paramagnetic states, at infinite temperature. Our objective is to train a neural network using this versatile catalog of ideal configurations, which spans the spectrum from extreme and ideal cases. In doing so, we enable the neural network to classify more complex scenarios through the process of generalization.

This is important because we will later use these trained networks to classify finite-temperature dilute systems. During training, the neural network never finds configurations with temperature and vacancies as input data. In this sense, our training is minimalist and lacks complex scenarios.

The minimalist training catalogue is composed by the following patterns: all spins pointing up or down, chessboard-like configurations, and horizontally or vertically alternating stripes. These configurations represent ideal orders at zero temperature when the spin couplings are positive (ferromagnetic), negative (antiferromagnetic), and different for each direction, respectively. Additionally, random configurations of spins are also employed, representing paramagnetic states at infinite temperature. A representative sample of the mentioned configurations is depicted in upper and lower panels of the Figure \ref{fig:catálogo-square} corresponding to the square and honeycomb lattices, respectively. 

\begin{figure*}[t!]
    \centering
    \includegraphics[width=0.9\textwidth]{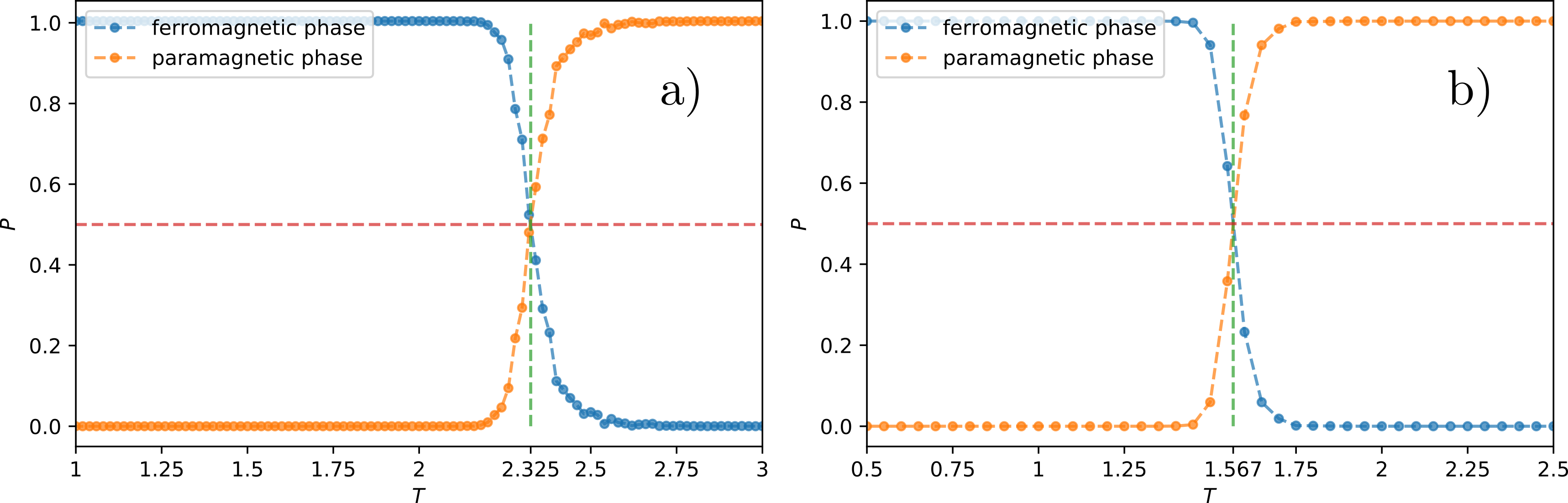}     
    \caption{Neural network's estimated probabilities (p) as a function of Temperature (T) corresponding to the pure Ising model ($\rho=1$) for: a) the square lattice with $N=40\times 40$ sites. b) the honeycomb lattice with $N= 2 \times 28 \times 28$ sites. The transition temperature is determined by the intersection of both probabilities, marked by a dashed vertical line. This procedure and notation are consistent throughout the paper.}
    \label{fig:prob1}
\end{figure*}

Concerning the training process and the neural network utilized, the fundamental principles of neural networks, encompassing their architecture and training methods, have become standard practices. Therefore, we focus on delineating the essential architecture of our specific neural network and the generation of training data. For a comprehensive understanding of this topic, we recommend referring to the cited references \cite{bengio2017deep,nielsen2015neural,aggarwal2023neural,géron2022hands,chollet2021deep}.

Our training approach employs a simple neural network design, aligning with the deliberate restriction of resources, both in the training dataset and in the architecture of the neural network itself.

We employed a two-layer fully-connected (DNN) neural network  \cite{bengio2017deep}. The first layer utilizes the \textit{ReLu} \cite{chollet2021deep} activation function and consists of $N$ neurons, where $N$ represents the number of sites within our spin lattice. The second layer utilizes the \textit{softmax} \cite{chollet2021deep} activation function with a neuron count equivalent to the number of orders for classification. Our choice of cost function is the \textit{cross-entropy} \cite{géron2022hands}, and for optimization, we utilize \textit{RMSProp} \cite{chollet2021deep} (Root Mean Squared Propagation). The implementation is carried out in \href{https://www.tensorflow.org/}{TensorFlow}.

In the training process, we generated datasets comprising 9000 synthetic configurations copies, sampling in equal proportion all the configurations of the catalog. Specifically, for the square lattice, we created synthetic datasets for lattice sizes of $L\times L$ with $L$ set to 30 (900 sites) and 40 (1600 sites). Likewise, for the honeycomb lattice, datasets were generated for lattices of fixed dimensions of $2\times L\times L$ sites, where $L$ was chosen as 21 (882 sites) and 28 (1568 sites).

During training, we utilized batches of 60 and ran for 5 epochs \cite{chollet2021deep}. This sufficed to attain an accuracy (the fraction of correct predictions) approaching 100$\%$, owing to the simplicity of the catalog.

\section{Results}
\label{results}

In this section, we assess the performance of our model, which has been trained using synthetic minimalist configurations, on the diluted Ising model at finite temperature and vacancy concentration, denoted as $\rho$. To generate configuration samples of the diluted Ising model at these conditions, Montecarlo simulations are employed, following the methodology outlined in reference \cite{Barkema}.

It is important to recall that these simulated configurations are not utilized for training the neural network; rather, they serve the purpose of validating and testing its generalization capabilities. From this perspective, the origin of the data, whether from simulations or experiments, becomes irrelevant. The neural network has already been trained on a curated dataset of ideal configurations and does not depend on these additional data for its training phase. However, it does rely on them for validation purposes.

For the sake of clarity and conciseness, we omit the technical details of the Montecarlo simulations performed to generate the validation samples for our neural network. These details are provided in the Appendix \ref{apendix-montecarlo} . In this section, we assume that these samples have been generated, and we proceed to describe their utilization.

\begin{figure*}[t!]
    \centering
    \includegraphics[width=0.9\textwidth]{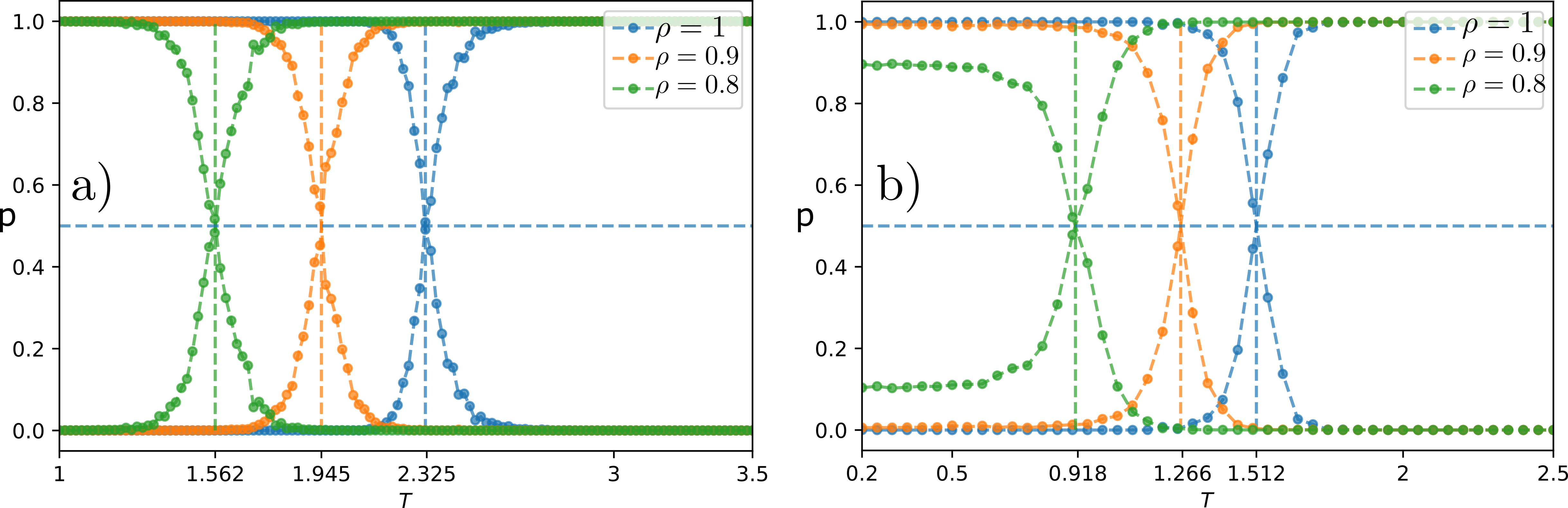}    
    \caption{ Neural network's estimated probabilities (p) as a function of Temperature (T) in the high density regime of the dilute Ising Model, corresponding to: a) The square lattice with $N=40\times 40$ sites. b) The honeycomb lattice with $N= 2 \times 28 \times 28$ sites. Curves correspond to $\rho=1$ (blue), $\rho=0.9$ (orange) and  $\rho=0.8$ (green). In both Figures, it can be observed that as the spin density of the lattice decreases, the critical temperature $T_c$ also decreases.
    }
    \label{pred_p01_p08}
\end{figure*}

\subsection{Critical temperatures on pure spin-lattices}

The generalisation process is evaluated in two stages. First, we analyse the pure Ising model. To do this, we present the neural network with simulated configurations of the model at different temperatures. 

At high temperatures (far from the transition), the model exhibits the characteristics of the paramagnetic state (disordered). Therefore, using the available catalogue of ideal configurations, the neural network easily selects the random configuration. The same applies to the ordered states at low temperatures (on the other side of the transition), where the model displays a ferromagnetic character. In this case, the neural network readily associates these states with the catalogue configurations where all spins point either up or down. In both situations, the neural network discards all other configurations in the catalogue that do not correspond to paramagnetic or ferromagnetic orders.

However, as the temperature approaches the transition point, whose value is analytically known in the thermodynamic limit, the network predictions become increasingly imprecise. Although it still assigns a larger probability to the random configuration above the transition and to the ferromagnetic configuration below it, its predictions become less reliable compared to the extreme temperatures.

To determine the transition using a neural network, we define the transition temperature as the value at which the prediction probabilities for the disordered and ordered phases are equal \cite{corte2021exploring}.  

To illustrate the approach, we present results for the square lattice and subsequently for the honeycomb lattice. 
In the Figure \ref{fig:prob1}-a, we show  the probability curves as a function of temperature $T$, which is hereafter expressed in units of $|J|/k_B$, for the Ising model on a square lattice of size $N=L \times L$, with $L=40$. 
For each temperature $T$, the probability is determined as an average of the predictions, belonging to 300 independent simulations. Only the prediction probabilities for the ferromagnetic and paramagnetic phases are represented, by blue and orange dots connected by dotted lines, respectively. The remaining catalog-trained phases (stripes and Néel) consistently yielded nearly zero probabilities and are consequently not displayed in the Figure. By intersecting the two curves at the point where both probabilities are equal, we estimate the critical temperature ($T_c \approx 2.325$). In multiclass classification, the predicted phase corresponds to the one with the highest probability. In this scenario with two dominant classes, the transition point is identified when both class probabilities reach 0.5, indicated by their intersection. Note that the critical temperature obtained through DNN classification is in good agreement with the analytical value ($T_c \simeq 2.269$). 

The same procedure performed for the square lattice was then carried out on the honeycomb lattice. The results are shown in Figure \ref{fig:prob1}-b. Again the neural network selects only the two intervening states, whose probability curves intersect at 
$T_c \approx 1.567$ for $L=28$. These results are in good agreement with the analytical solution, $T_c \simeq 1.519$.

It's worth recall that the neural network was never trained on the temperature variable and only received extreme ideal ordered and disordered configurations. Consequently, we could only anticipate qualitative agreement between the analytically determined transition temperature and the network's determination, particularly considering the finite size of the validation samples. However, the DNN identifies the transition with substantial quantitative agreement. In essence, the order parameters derived from the ideal configurations are sufficiently robust to predict the transition, even when it falls far from the extremes, exhibiting distinct properties compared to those of the ordered and disordered phases.

\subsection{Dilute spin-lattices}

After establishing the neural network's correct classification of magnetic phases in the Ising model on square and honeycomb lattices, we seek to test its generalization capabilities further. Specifically, we aim to determine whether the neural network, trained on minimalist data, can predict changes in the critical temperature resulting from modifications in the spin lattice. One such alteration includes modifying the critical temperature by varying the spin density, achieved through the introduction of vacancies (holes).

Furthermore, we investigate whether the neural network can effectively classify the phases of the diluted model, even in the absence of explicit training with diluted spin configuration data.

In this section, we delve into the neural network's classification phases within square and honeycomb lattices, characterized by different spin densities ($\rho$) obtained from the Montecarlo simulations. These classifications will allow us to estimate critical temperatures $T_c(\rho)$ as a function of spin density.

\subsubsection{High-density regime}

We begin our exploration by examining the magnetic phases within the low(high) hole(spin)-density regime ($\rho \approx 1$), which is significantly distant from the percolation density. Just as in the pure case, the spin configurations were extracted from Montecarlo simulations. To construct the predicted probability curves, we computed the temperature ($T$) predictions by averaging results from 300 simulations.

\begin{figure*} [t!]
    \centering
    \includegraphics[width=1\textwidth]{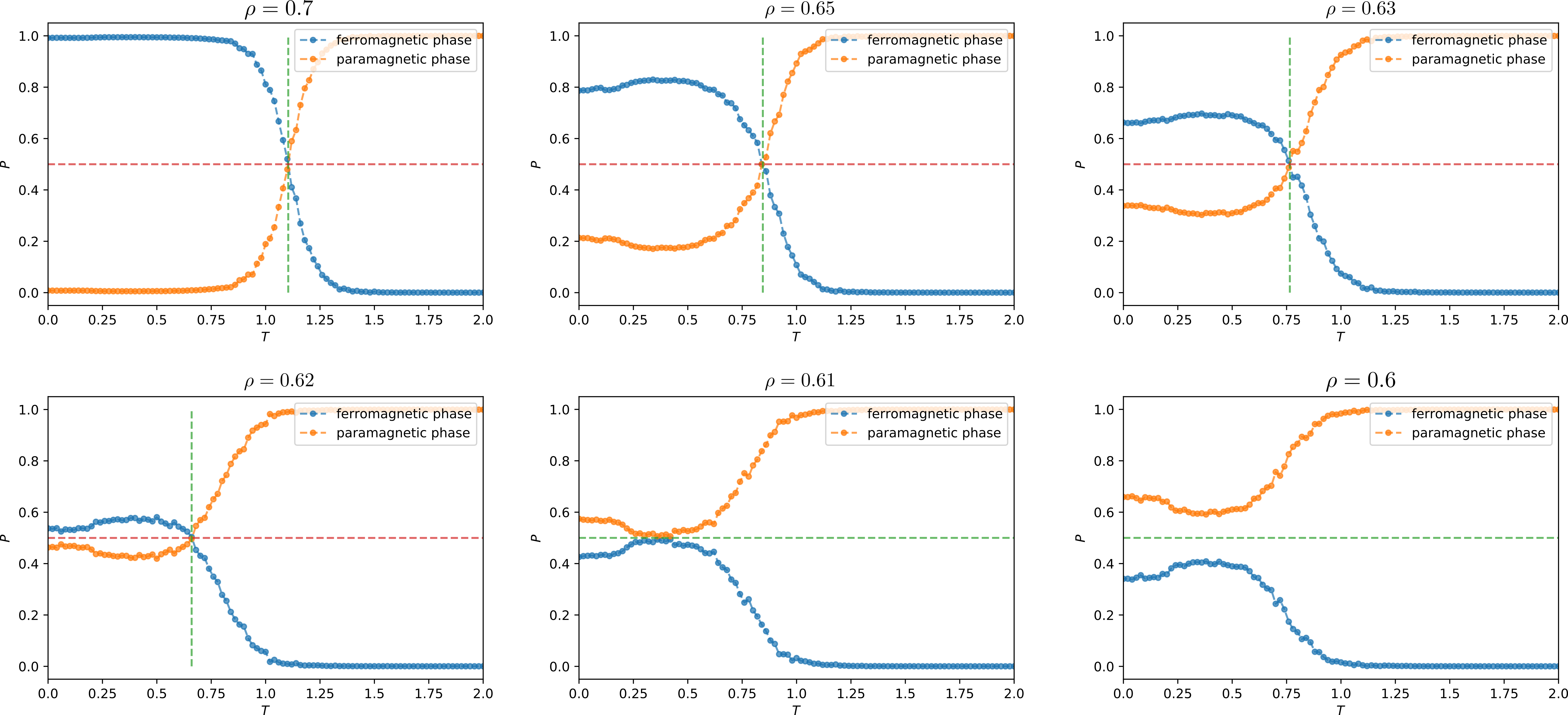}    
    \caption{Phase classification results results for the dilute Ising model on the square spin lattice with $L=40$, in the low-density regime, across densities from $\rho=0.7$ to $\rho=0.6$. Phase transitions are observed at the higher densities, signaled by probability curve intersections. However, from $\rho = 0.61$, no phase transition is observed, indicating a disordered state with $T_c=0$. This suggests a percolation density for this lattice size around this value, close to the approximate theoretical value.
    }
    \label{pred_L40_low_densities}
\end{figure*}

The Figure \ref{pred_p01_p08} illustrates the predicted probability curves, corresponding to ferromagnetic and paramagnetic configurations, for both the square lattice (left panel) and the honeycomb lattice (right panel) across three different densities: $\rho=0.8$ (green), $\rho=0.9$ (orange), and $\rho=1$ (blue). These curves show a noticeable trend where, as $\rho$ decreases, the critical temperature $T_c$ also decreases.

It is worth mentioning that in the honeycomb lattice with $\rho=0.8$ (as depicted by the green curves in the right panel of Figure \ref{pred_p01_p08}) below the critical temperature $T_c$, the probability for the ferromagnetic phase falls below 1. This observation indicates our proximity to the percolation density. This phenomenon, where the prediction probability for the ferromagnetic phase diminishes as we approach the percolation density, becomes more pronounced at lower densities.

\subsubsection{Low-density regime and percolation}

So far we have shown that the neural network can correctly classify the magnetic states with densities $\rho \approx 1$, even though the network was trained solely on configurations without vacancies nor temperature.  We will now shift our focus to examining the classification of low spin densities with the aim of detecting the percolation density $\rho_c$. An indicator that the percolation density has been reached is that the system becomes disordered at any finite temperature, i.e. $T_c=0$ \cite{stauffer}. In other words below the percolation density, the system does not exhibit phase transition \cite{martins2007}. Consequently, in the probability curves, we should not observe any crossovers when $\rho < \rho_c$.

For the square lattice, it is established that the percolation density is $\rho_c \approx 0.5927$ \cite{p_c}. This reference provides a basis for comparing the neural network results when examining densities in proximity to the percolation density.

In Figure \ref{pred_L40_low_densities}, we present the outcomes derived from the neural network classification for the square spin lattice with a characteristic length of $L=40$. These results pertain to decreasing densities, namely, $\rho=0.7$, $\rho=0.65$, $\rho=0.63$, $\rho=0.62$, $\rho=0.61$ and $\rho=0.6$.

There are several observations to highlight. In the case of density $\rho=0.7$, a clear phase transition occurs, as evidenced by the intersection of the probability curves for the two phases at $T_c \approx 1.10$. Similarly, for densities $\rho=0.65$, $\rho=0.63$, and $\rho=0.62$, phase transitions are also evident due to the intersection of the probability curves, corresponding to each phase, at temperatures $T_c \approx 0.850$, $T_c \approx 0.77$, and $T_c \approx 0.67$, respectively. For these densities it can be observed that, as $T$ decreases, the prediction probability curve for the ferromagnetic phase (blue curves) no longer remains at 1, as it was the case for density $\rho=0.7$ and higher. Conversely, the prediction probability curve for the paramagnetic phase (orange curves) no longer remains at 0. 

On the other hand, for densities $\rho=0.61$ and $\rho=0.6$, the phase transition is no longer observed as there is no intersection between the probability curves corresponding to paramagnetic and ferromagnetic phases. This implies that the percolation density $\rho_c$ has been surpassed. Notably, at all temperatures, the system remains disordered, with the prediction probability for the paramagnetic phase being predominant, i.e. $T_c = 0$. 

Based on the analysis of the square lattice with a characteristic length of $L=40$, it can be inferred that the percolation density $\rho_c$ is upper-bounded by $\rho \approx 0.61$ for this size, in close proximity to the theoretical value. \\

\begin{figure*} [t!]
    \centering
    \includegraphics[width=1\textwidth]{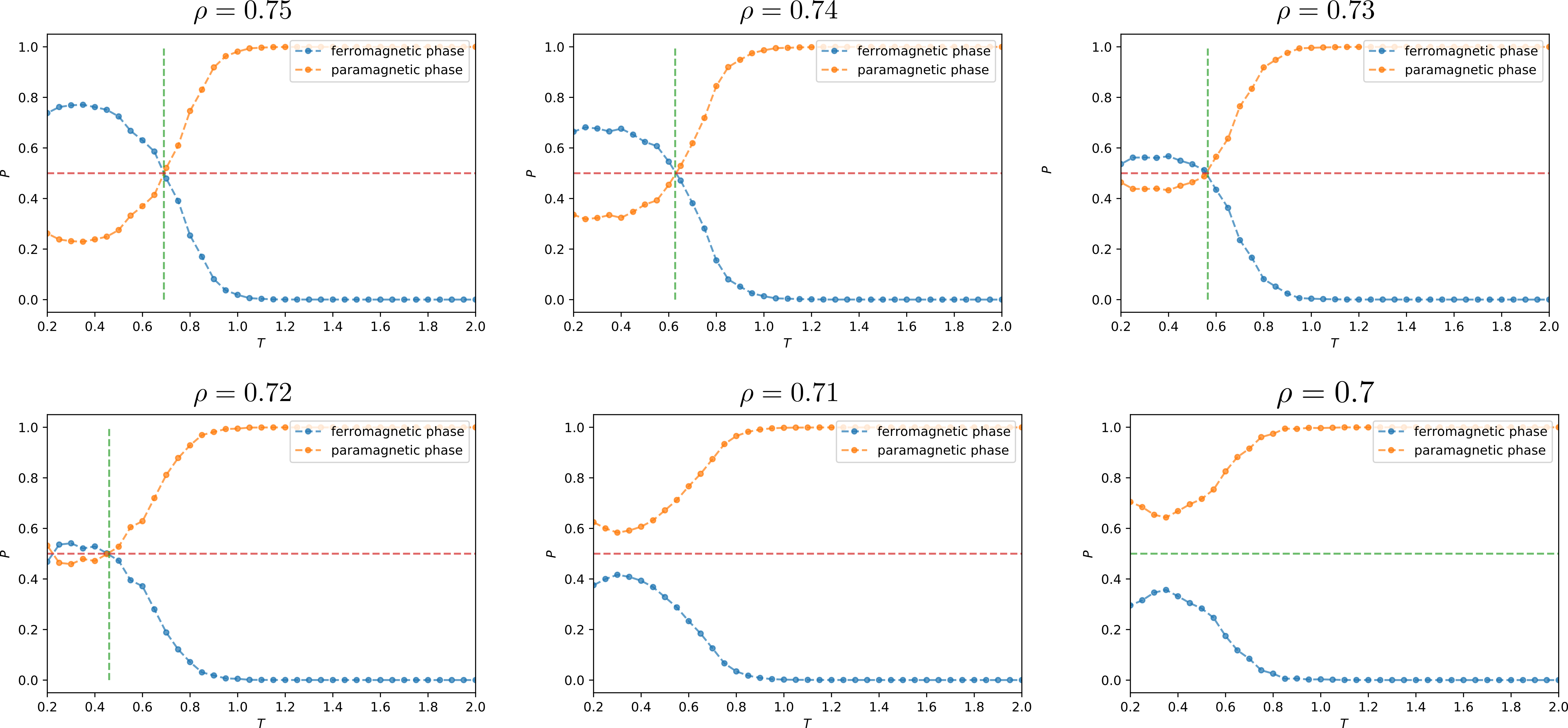}
    \caption{Phase classification results for the dilute Ising model on a $L=28$ honeycomb lattice in the low density regime. Phase transitions are observable up to $\rho=0.72$, marked by probability curve intersections. Notably, higher prediction uncertainty is evident, attributed to the honeycomb lattice's higher percolation density compared to the square lattice. Beyond $\rho=0.71$, no further probability curve intersections occur, establishing an upper percolation density limit for this size, near the theoretical value.}
    \label{pred_HL21_densidades bajas}
\end{figure*}

The same analysis as described previously was carried out for the honeycomb lattice, with the estimated percolation density from the literature being $\rho_c \approx 0.6962$ \cite{p_cH}. In this case, we considered densities proximate to the percolation density, specifically $\rho = 0.75$, $\rho = 0.74$, $\rho = 0.73$, $\rho = 0.72$, $\rho = 0.71$ and $\rho = 0.7$.

The Figure \ref{pred_HL21_densidades bajas} presents the results, akin to the square lattice, for a characteristic size $L=28$. In this case, a phase transition is observable up to $\rho \approx  0.72,$ characterized by the intersection of probability curves. However, even from the first panel, a more pronounced prediction uncertainty becomes evident, attributed to the higher percolation density in the honeycomb lattice compared to the square lattice. Beyond $\rho_c \approx 0.71$, no further intersections of probability curves are observed, thereby establishing an upper limit on the percolation density for this lattice size, albeit in close proximity to the theoretical value.

\subsection{Phase diagram}

In the preceding section, we determined the critical temperatures ($T_c$) for several spin concentrations ($\rho$) using neural network classification. To facilitate comparison with alternative approaches, we plotted these $T_c(\rho)$ curves for each spin lattice configuration previously obtained. Subsequently, we performed curve fitting \cite{press2007numerical-chap15}, employing a nonlinear model akin to the one found in existing literature \cite{selke, Neda}. The model proposed is expressed as follows
\begin{equation}
    \frac{T_c(\rho)}{T_c(1)}= \frac{K}{\log(a-b(1-\rho))}.
    \label{ajuste}
\end{equation}
Here, $K$, $a$, and $b$ represent non-universal parameters subject to fitting, while $\rho$ denotes the density of the spin lattice corresponding to the critical temperature $T_c$. The phase transition curve separates the ordered phase, located below the curve, from the disordered phase, positioned above it, for each type of spin lattice.

\begin{figure*} [t!]
    \centering
    \includegraphics[width=1\textwidth]{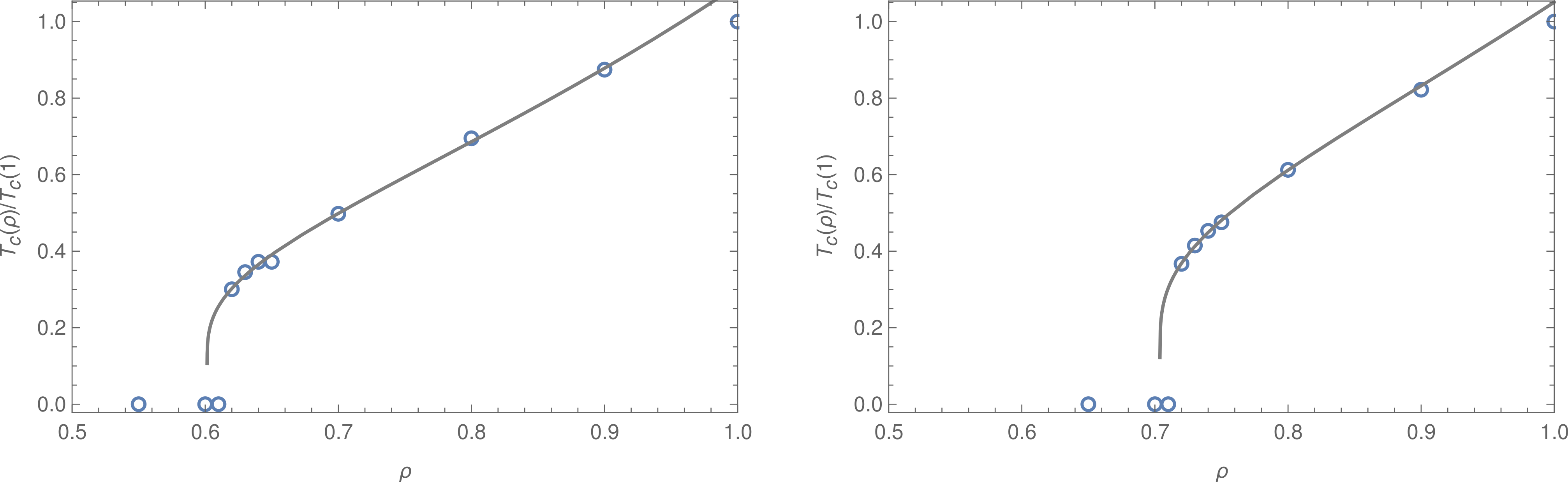}
    \caption{Left: Normalized critical temperature curve $T_c$ predicted by the neural network (blue open circles) as a function of spin density $\rho$ for diluted square lattice of $L \times L$ sites with $L=40$. The Figure also displays the fit (blue solid line) performed based on the proposed model \ref{ajuste}. The ferromagnetic phase lies below the curve, while the paramagnetic phase resides above it. Note that the fitted curve does not intersect the $\rho$-axis due to the logarithmic divergence in the transition.  Right: The same results as in the left panel, but for the diluted honeycomb lattice with a specific size of $L = 28$, consisting of $2 \times L \times L$ sites.
    }
    \label{Tc_vs_p}
\end{figure*}

Figure \ref{Tc_vs_p} (left panel) shows curve fitting results for $T_c(\rho)$ of the $L \times L$ square lattice ($L=40$). The estimated percolation density is $\rho_c=$0.601, obtained directly from the logarithm divergence condition, i.e. $a-b(1-\rho_c) = 0$ in \ref{ajuste}, indicating $T_c$ approaching zero.

While the primary focus here is not to provide an exhaustive analysis of how the results scale with size, the table \ref{tabla Tc_vs_p cuadrada} presents the fitting outcomes and transition temperatures on the square lattice for both $L=30$ and $L=40$.

\begin{table} [H]
\centering
\begin{tabular}{cllll}
\hline
\multicolumn{1}{|c|}{Square lattice of $L\times L$ sites}     & \multicolumn{1}{c|}{$K$}      & \multicolumn{1}{c|}{$a$}     & \multicolumn{1}{c|}{$b$}  & \multicolumn{1}{c|}{$\rho_c$}      \\ \hline
\multicolumn{1}{|c|}{$L=30$}  & \multicolumn{1}{c|}{-1.405} & \multicolumn{1}{c|}{0.253}  & \multicolumn{1}{c|}{0.653}
& \multicolumn{1}{c|}{0.612}\\ \hline
\multicolumn{1}{|c|}{$L=40$}  & \multicolumn{1}{c|}{-1.279} & \multicolumn{1}{c|}{0.311}  & \multicolumn{1}{c|}{0.780}  & 
\multicolumn{1}{c|}{0.601}  \\ \hline

       &                              &                              &  
       & \\ 
\end{tabular}
\caption{Fitted parameters from the function (\ref{ajuste}) and their respective percolation densities $\rho_c$, for the diluted square lattices with $L\times L$ sites ($L=30$ and $L =40$).} 
\label{tabla Tc_vs_p cuadrada}
\end{table}

In the case of the square lattice, as we have mentioned, the percolation density is known to be approximately $\rho_c  \approx 0.59274621(13)$ \cite{p_c}. Remarkably, the fitted value of $\rho_c$ from our data aligns closely with this literature value. This agreement is noteworthy, given that our network was trained under minimalist conditions, without vacancies or temperature considerations, and without utilizing simulation data, all of which are far from the percolation regime.

Following the same procedure carried out for the square lattice, Figure \ref{Tc_vs_p} (right panel) shows the curve fit for the $T_c(\rho)$ for the $2 \times L \times L$ sites honeycomb lattice, with $L=28$, where the estimated percolation density is $\rho_c \approx 0.703$. Additionally, Table \ref{tabla Tc_vs_p honeycomb} provides the fitted parameters and critical density estimates for two lattice sizes.

Given that the established percolation density for the honeycomb network is $\rho_c \approx 0.6962(6)$ \cite{p_cH}, the value obtained through the fitting process for $\rho_c$ is very satisfactory, for analogous reasons as those outlined for the square lattice.

\begin{table} [H]
\centering
\begin{tabular}{cllll}
\hline
\multicolumn{1}{|c|}{Honeycomb lattice of $2 \times L\times L$ sites}     & \multicolumn{1}{c|}{$K$}  & \multicolumn{1}{c|}{$a$}     & \multicolumn{1}{c|}{$b$}  & \multicolumn{1}{c|}{$\rho_c$}      \\ \hline
\multicolumn{1}{|c|}{$L=21$}  & \multicolumn{1}{c|}{-1.623} & \multicolumn{1}{c|}{0.223}  & \multicolumn{1}{c|}{0.781}
& \multicolumn{1}{c|}{0.714}\\ \hline
\multicolumn{1}{|c|}{$L=28$}  & \multicolumn{1}{c|}{-1.647} & \multicolumn{1}{c|}{0.208}  & \multicolumn{1}{c|}{0.704}  & 
\multicolumn{1}{c|}{0.703}  \\ \hline

       &                              &                              &  
       & \\ 
\end{tabular}
\caption{Adjusted parameters derived from equation (\ref{ajuste}) alongside their corresponding percolation densities $\rho_c$ for the diluted honeycomb lattices with $2\times L \times L$ sites ($L=21$ and $L =28$).}
\label{tabla Tc_vs_p honeycomb}
\end{table}

\section{Conclusions}
\label{conclusions}

We investigated the order-disorder transition in dilute Ising models through neural networks trained using a `minimalist' approach.

The training process involved utilizing a basic catalog of configurations that encompassed ideal ordered states at $T=0$, along with random configurations representing the infinite-temperature paramagnetic phase. This allowed the neural network to be trained on a variety of ideal configurations, enabling it to effectively classify these synthetic samples.

By exploiting the generalization ability of the previously trained neural networks, they were used to classify configurations obtained from Montecarlo simulations corresponding to dilute Ising models. Since the neural network was trained with a set of ideal configurations and does not rely on specific data for training, this methodology allows the data to come from simulations or experiments, without any difference.

Initially, we employed the trained neural network to categorize phases within a pure Ising model. The minimalist neural network was provided with configurations derived from simulations of spin models on both square and honeycomb lattices at various temperatures, resulting in probability curves for each magnetic order. The neural network, trained on very simple configurations, demonstrated the capacity to generalize and accurately classify the ordered and disordered phases.

Subsequently, we introduced vacancies into the spin lattices to increase the complexity of the model. Remarkably, the minimalist neural network displayed the capability to generalize and effectively classify the different configurations, even though it was never trained on spin lattices featuring vacancies. The neural network exhibited its ability to generalize from ideal spin lattice configurations to scenarios involving finite temperatures and diluted spin lattices.

Analysis of the temperature-dilution phase diagram of the dilute Ising model, spanning paramagnetic and ferromagnetic phases, showed that the transition curve obtained via the neural network is consistent with the expected approximate analytical predictions. This illustrates the power of minimalist neural networks to generalize and classify appropriately the different phases of magnetic orders, for different values of temperature and dilution. 

The percolation densities $\rho_c$ for the spin lattices were indirectly obtained through fitting the data from the $T_c(\rho)$ curves. These $\rho_c$ values, derived from the classifications performed by the minimalist neural networks, exhibit a good quantitative alignment with those reported by other approximate methods. This holds true for both the square lattice and the honeycomb lattice.

The findings presented in this paper suggest that minimalist neural networks offer an efficient data analysis approach that does not necessitate extensive resources, making it a cost-effective choice. Despite the synthetic training data, the neural networks yield highly satisfactory results for the examination of dilute spin lattices. This assessment is further reinforced when considering the trade-off between simplicity and complexity in comparison to conventional methods like Montecarlo simulations.

In summary, this study provides a framework for exploring the capabilities of minimalist networks in analyzing and categorizing complex spin lattice configurations, including those featuring frustration and other ingredients. Additionally, due to their pre-trainability with a expandable catalog and ease of deployment, these networks hold potential as valuable predictive tools for both synthetic and experimental data.

This approach has the potential to open new perspectives in the study of complex multicomponent systems in physics and other fields of science with minimalist artificial intelligence tools.

\section{acknowledgments}

Carlos A. Lamas and Marcelo Arlego acknowledge support from CONICET. This work was partially
supported by CONICET, Argentina (grant no PIP 2332) and UNLP
(grant 11X987).

\appendix

\section{Montecarlo simulations}\label{apendix-montecarlo}
\label{appendix}

In this section, we provide additional information regarding the Montecarlo simulations conducted in this study. These simulations were performed to generate data samples exclusively for the purpose of validating the generality, rather than training, the neural networks employed.

All Montecarlo simulations were implemented in C language. These simulations were carried out for the ferromagnetic Ising model $(J>0)$ on both square and honeycomb lattices, employing the Metropolis-Hastings algorithm. We executed 300 independent simulations for each density $\rho$. 

In all simulations, periodic boundary conditions were enforced. The simulations started at a temperature above $T_c$, representing a disordered phase, and gradually reduced the temperature in intervals of $\Delta T$ until reaching temperatures below $T_c$, ultimately attaining the ordered phase. At each temperature point, both the temperature value and spin configuration were stored once equilibrium was achieved.

For the square lattice, we investigated two distinct system sizes: one comprising 900 sites (with a characteristic length of $L=30$) and the other consisting of 1600 sites ($L=40$). In all cases, periodic boundary conditions were employed. For both sizes, we conducted 300 independent simulations spanning the temperature range $T=[5, 0.2]$, with intervals of $\Delta T=0.02$ (resulting in 250 temperature values).

We monitored the systems over $50000 \times L^{2}$ Montecarlo steps after completing $20000\times L^{2}$ relaxation iterations. From each simulation, we extracted 250 spin configurations corresponding to each temperature value $T$.

\begin{figure} [t!]
    \centering
    \includegraphics[width=0.9\columnwidth]{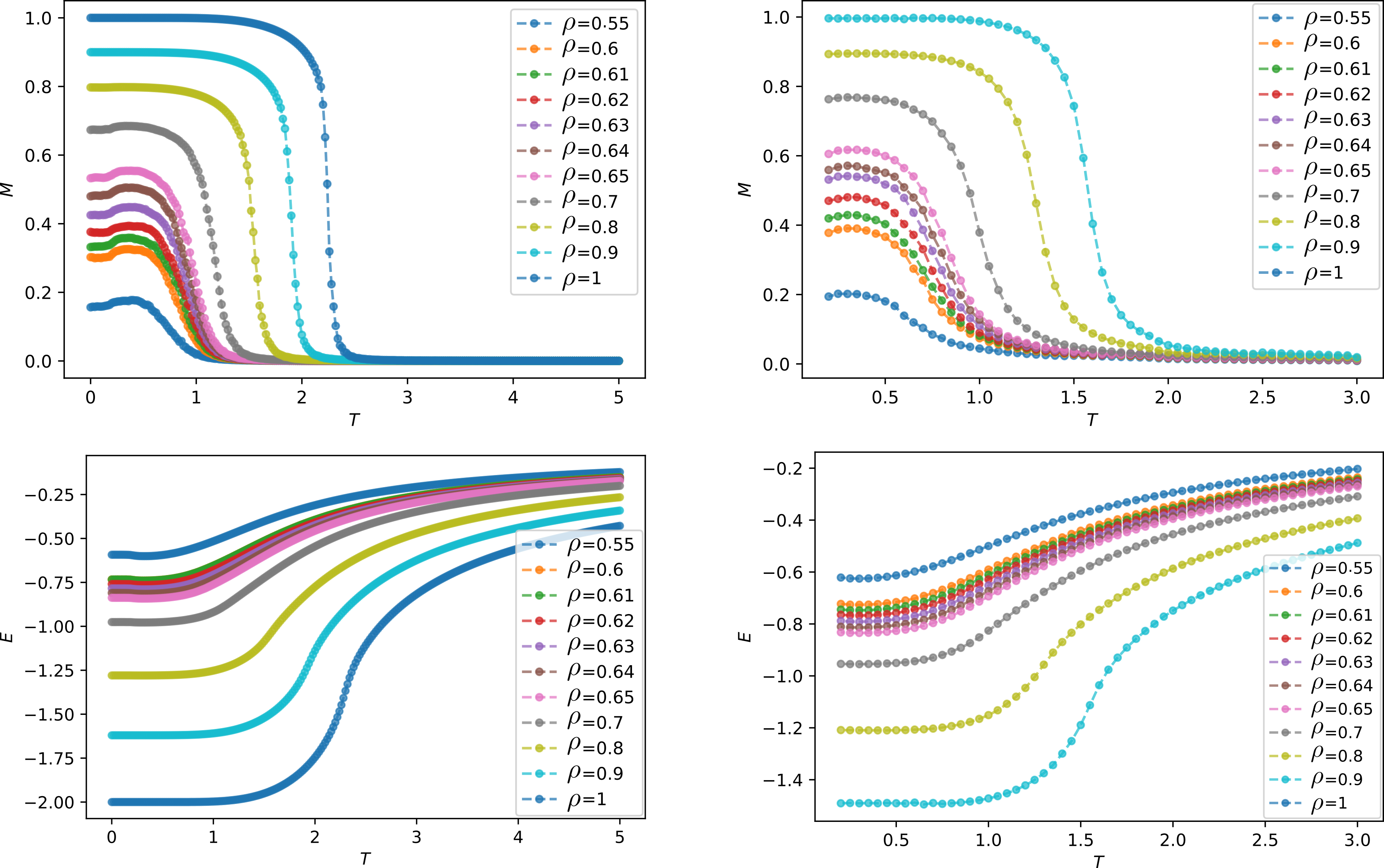}
    \caption{    
    Magnetization and energy per site plotted against temperature for the dilute Ising model on the square lattice (left column) with $L=40$ and the honeycomb lattice with $L=28$ (right column), for different spin densities.
    }
    \label{E_vs_T y M_vs_T}
\end{figure}

Considering the percolation density of the square lattice as $\rho_c \approx 0.5927$ \cite{p_c}, we performed simulations for the dilute Ising model, covering a range of densities, namely $\rho=1$ (representing the pure case), 0.9, 0.8, 0.7, 0.65, 0.64, 0.63, 0.62, 0.61, 0.6, and 0.55. In each instance, vacancies were randomly introduced into the lattice using a uniform distribution, until the desired spin density $\rho$ was achieved.

We followed a similar procedure for the honeycomb lattice. In this case, we considered system sizes of $2 \times L \times L$ sites. Specifically, we investigated two system sizes: one comprised of 882 sites ($L=21$), and the other consisting of 1568 sites ($L=28$). For both sizes, periodic boundary conditions were applied. In each case, we conducted 300 independent simulations covering the temperature range $T=[3, 0.2]$, with intervals of $\Delta T =0.05$ (resulting in 60 temperature values $T$). We monitored the systems for $290 \times L^2$ Montecarlo steps after completing $32  \times L^2$ relaxation iterations. From each simulation, we obtained 60 spin configurations corresponding to each value of $T$.

Considering the percolation density as $\rho_c  \approx 0.6962$ \cite{p_cH}, we conducted simulations with densities $\rho=1$ (representing the pure case), 0.9, 0.8, 0.74, 0.73, 0.72, 0.71, 0.7, and 0.65. Similar to the square lattice case, vacancies were randomly introduced into the lattice using a uniform distribution.

To visually depict the outcomes of our simulations, Figure \ref{E_vs_T y M_vs_T} shows graphical representations of magnetization and energy per site as functions of temperature for the dilute Ising model on the square lattice (left column) with $L=40$ and the honeycomb lattice with $L=28$ (right column) for the different spin densities analyzed.

The graphs presented in Figure \ref{E_vs_T y M_vs_T} exhibit the expected behavior. In the case of magnetization, it is evident that as the density $\rho$ decreases, the critical temperature $T_c$ also decreases, resulting in a reduction of the maximum magnetization per site.

Regarding the energy plots, as temperature ($T$) approaches zero, it is noteworthy that for the pure cases of the square lattice and honeycomb lattice, the energy attains its anticipated minima of -2 and -3/2, respectively.

\begin{figure}[t!]
    \centering
    \includegraphics[width=0.95\columnwidth]{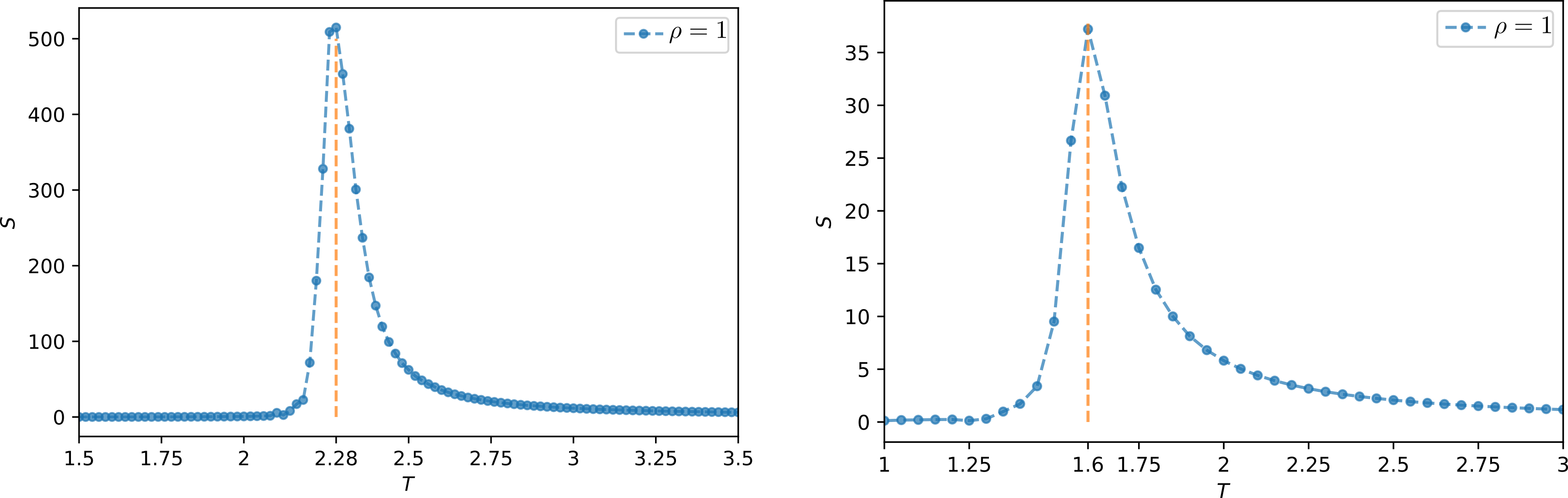}
    \caption{    
    Susceptibility versus temperature for the pure Ising model corresponding to the square lattice with $L=40$ (left) and the honeycomb lattice with $L=28$ (right).
    The data was obtained by averaging results from 300 independent Montecarlo simulations.  
    }
    \label{Tc_S}
\end{figure}

In addition, we investigated the susceptibility as a function of temperature for the pure Ising model ($\rho=1$). The Figure \ref{Tc_S} depicts the results for the square lattice with $L=40$ (left panel) and the honeycomb lattice with $L=28$ (right panel). These plots show that for the square lattice, the maximum occurs at $T_c \approx 2.28$, while for the honeycomb lattice, it occurs at $T_c \approx 1.6$. These results align with the analytically reported literature values, although it's important to note that these analytical values are derived in the thermodynamic limit, in contrast to our finite-size results.

Based on these observations, we conclude that the simulations are well-suited for extracting spin configurations at various temperature values, confirming their adequacy and suitability for our study.

\bibliographystyle{unsrt}
\bibliography{refs}

\end{document}